\documentclass[reprint,amsmath,amssymb,aps,pra,superscriptaddress,nofootinbib,floatfix
]{revtex4-2}

\usepackage{graphicx}
\usepackage{dcolumn}
\usepackage{bm}
\usepackage{hyperref}
\usepackage{amsmath}

\usepackage[
    protrusion=true,
    activate={true,nocompatibility},
    final,
    tracking=true,
    kerning=true,
    spacing=true,
    spacing=nonfrench,
    factor=1000]{microtype}
\hypersetup{
 colorlinks=true, 
 linkcolor=blue, 
 citecolor=blue,        
 filecolor=blue,      
 urlcolor=blue}
 \usepackage{xcolor}
 \usepackage{ulem}

\newcommand{\pypx}[2]{\frac{\partial #1}{\partial #2}}

\begin{document}

\preprint{APS/123-QED}


\title{Frequency comb stability, bistability, and self-starting in normal dispersion microresonators}

\author{Logan~Courtright}
\email{lcourt1@umbc.edu}
\affiliation{University of Maryland Baltimore County, Baltimore, MD, 21250, USA}
\author{Alexey~Sukhinin}%
\affiliation{University of Maryland Baltimore County, Baltimore, MD, 21250, USA}
\author{Pradyoth~Shandilya}%
\affiliation{University of Maryland Baltimore County, Baltimore, MD, 21250, USA}
\author{Tanvir~Mahmood}%
\affiliation{US Army Research Laboratory, 2800 Powder Mill Road, Adelphi, MD, 20783, USA}
\author{James P.~Cahill}%
\affiliation{US Army Research Laboratory, 2800 Powder Mill Road, Adelphi, MD, 20783, USA}
\author{Curtis R.~Menyuk}%
\affiliation{University of Maryland Baltimore County, Baltimore, MD, 21250, USA}


%
%
%

\begin{abstract}
Recent work has demonstrated that it is possible to generate frequency combs in normal dispersion microresonators by taking advantage of fabricated mode coupling that induces a mode split. In particular, photonic crystal resonators periodically modulate the microresonator's width, which geometrically creates mode coupling. In contrast to frequency combs generated by microresonators in the anomalous dispersion regime, which are typically generated using bright solitons, the frequency combs generated by normal dispersion regime microresonators are dark pulses that are commonly referred to as dark solitons or platicons. In this work, we examine the linear stability of platicons in the pump-detuning and pump-power parameter space as the mode split strength and cavity dispersion is varied. Our work reveals a broad region of stability for platicons. Our work also reveals the existence of fingers at high power at whose boundaries a sharp transition occurs and inside of which platicon solutions are bistable. Finally, we investigate the overlap of the region where platicons are stable with the region where continuous waves are modulationally unstable, and it is possible for a platicon to self-start. Simulations show that platicons consistently self-start inside this region.
\end{abstract}

\maketitle


\section{Introduction}

One of the most consequential inventions of the early 21-st century has been the self-referenced, or $f$-$2f$-locked, optical frequency comb (OFC) \cite{jones_carrierenvelope_2000}. This invention has impacted nearly every discipline of physics, from the most fundamental to the most applied. Within the first decade, many potential applications were laid out \cite{diddams_evolving_2010} and since that time, many of them have been realized in practice \cite{diddams_optical_2020}. Most OFCs to date have been generated using passively modelocked lasers, which were first discovered in the 1960s \cite{mocker_mode_1965,demaria_self_1966}. In these lasers the cavity modes self-organize into pulses, known as dissipative solitons \cite{grelu_dissipative_2012}, that circulate periodically and, absent noise, produce an evenly spaced comb of frequencies. Some mode-locked lasers require active triggering, but many self-start, organizing spontaneously from noise \cite{keller_femtosecond_1991,haus_selfstarting_1991}. From an applications standpoint, this property is desirable since it simplifies the laser design. From a more fundamental standpoint, self-starting is a remarkable example of symmetry-breaking whose close connection to the thermodynamics of crystallization has been studied in \cite{gordon_phase_2002,weill_critical_2005,gordon_selfstarting_2006}. It has long been known that the necessary condition for the system to self-start is for continuous waves (CWs) to be unstable in the same parameter regime where the comb-producing waveform is stable \cite{chen_selfstarting_1995,haus_selfstarting_1991}.

In 2007, it was discovered that OFCs can also be generated in microresonators \cite{delhaye_optical_2007}. Since then, a large amount of work has been done to explore the potential of microresonator devices and to integrate them into chip-scale devices alongside other photonic components \cite{kippenberg_microresonatorbased_2011, gaeta_photonicchipbased_2019,chang_integrated_2022}. Microresonator OFCs (microcombs) offer the prospect of reducing the size, weight, power, and cost (SWAP-C) relative to other OFC technologies. These devices are also a remarkable laboratory for nonlinear dynamics, supporting a wide diversity of bright waveforms that can be obtained in the anomalous dispersion regime, including bright solitons \cite{herr_temporal_2014}, soliton molecules \cite{wang_universal_2017}, Turing rolls \cite{qi_dissipative_2019}, soliton crystals \cite{cole_soliton_2017, karpov_dynamics_2019}, and soliton liquids and glasses \cite{cole_soliton_2017, young_soliton_2026}, as well as dark waveforms that can be obtained in the normal dispersion regime, such as dark solitons and platicons \cite{godey_stability_2014,parra-rivas_dark_2016,parra-rivas_origin_2016,lobanov_frequency_2015}. However, microcombs have historically been difficult to access due to thermal instabilities \cite{carmon_dynamical_2004,leshem_thermal_2021} and competition with continuous waves that are stable in the same parameter regime \cite{godey_stability_2014}. Consequently, microcombs do not typically self-start, and reports of self-starting to date are limited \cite{lihachev_platicon_2022,he_selfstarting_2019}.

Several techniques have been developed to make it easier to access microcombs, including induced mode coupling, either by coupling multiple resonators together \cite{miller_tunable_2015,kim_turnkey_2019}, injection-locking a laser to the resonator \cite{liang_generation_2014,lihachev_platicon_2022}, or by sinusoidally modulating the sidewall of a single resonator \cite{yu_spontaneous_2021,yu_continuum_2022}. While mode coupling makes it easier to access bright waveforms and increases the efficiency with which they are generated \cite{helgason_surpassing_2023,liu_breaking_2025a}, mode coupling appears to be essential for the generation of dark waveforms; they have only been obtained to date with mode coupling. Despite this limitation, dark waveforms have the advantage that they can be created in the normal dispersion regime, while bright waveforms must be created in the anomalous dispersion regime, which leads to designs that are challenging to fabricate with standard materials like silicon nitride \cite{pfeiffer_photonic_2016, liu_highyield_2021}.

Given the rich diversity of OFC-producing waveforms that exist in microresonators, it is not surprising that bistable and even multi-stable solutions can co-exist in the same parameter regime and can exhibit hysteresis \cite{godey_stability_2014}. Bistability and hysteresis are generic features of driven nonlinear systems, arising in contexts ranging from semiconductor tunnel diodes \cite{esaki_new_1958} to driven nanomechanical resonators \cite{aldridge_noiseenabled_2005}, and, beyond their fundamental interest, are useful for switching \cite{katine_currentdriven_2000} and random-number generation \cite{marandi_alloptical_2012}. However, this phenomenon is a hindrance to microcomb generation. A broad parameter region with limited competition from other waveforms is needed, which is one reason for the strong experimental focus to date on single bright solitons. They occupy a broad region in the experimental parameter space where they compete with only CWs \cite{barashenkov_existence_1996,qi_dissipative_2019}.

The key theoretical question, from an experimental standpoint, is where stable microcomb-producing waveforms exist in the experimentally accessible parameter space and how to access them. Snaking continuation methods that generate families of stable and unstable waveforms are a powerful approach to find a multiplicity of waveforms and have been used to study both bright \cite{parra-rivas_dynamics_2014} and dark \cite{parra-rivas_dark_2016,parra-rivas_origin_2016} microresonator waveforms. However, this approach does not determine the boundaries of a stable region in the parameter space where a waveform of a particular type exists and can be continually deformed as parameters change without bifurcating into a solution of a different type. Simulations of the evolution equations, a common approach in experimental papers, can only simulate solutions over a tiny fraction of the experimentally relevant timescales and is ambiguous near the boundary of a stable region. A third approach is to use boundary-tracking methods that continue the solution along several paths that are parallel to the boundary and use the spectrum of the linearized evolution equation to extrapolate its location \cite{wang_boundary_2014,qi_dissipative_2019,courtright_automatically_2021}. A difficulty with this approach in principle is that it is possible to miss an internal island of instability, although to date no such internal islands have been reported.

In this paper, we describe a theoretical and computational study of dark waveforms that we refer to as platicons, following Lobanov et al. \cite{lobanov_frequency_2015}. The work that we will present unites all the themes that we have discussed. We consider the solutions of a modified Lugiato-Lefever equation (LLE) in the four-dimensional parameter space that consists of the pump amplitude, the pump frequency detuning, the chromatic dispersion, and the mode coupling. We begin by using boundary tracking to determine the stability boundary of the platicon solutions as the mode coupling increases. In contrast to the family of cnoidal waves that consists of Turing rolls and perfect soliton crystals, where each periodicity has its own stable region \cite{qi_dissipative_2019}, the stability region of the platicons is a single, broad region regardless of the number of phase oscillations in the dark portion of the platicons. As the mode coupling increases, we find that the stable region moves towards and then encompasses zero detuning, which implies greater coupling efficiency. We also find that at large detuning, there are internal stability boundaries that create ``fingers'' in which two stability boundaries intersect and then disappear, demonstrating that internal stability boundaries can exist. Within the broad stable region, at detunings below where the fingers appear, the platicon deforms smoothly as system parameters change, indicating it is robust, which is useful for microcomb applications. We next examine the behavior of the solutions as the system passes into and out of the fingers. We find that the solutions are bistable and exhibit hysteresis so that they only make a sharp transition when passing in one direction through a finger boundary.

Finally, we examine the conditions under which platicons can self-start. To self-start, it is necessary that the stable region for the platicons have a large overlap with the region where CWs are unstable. Yang et al. \cite{yang_modulational_2025} previously calculated the CW instability region when mode coupling is present, using a procedure like that of Godey et al. \cite{godey_stability_2014}, and they showed numerically that platicons can self-start where CWs are unstable. Here, we systematically examine the overlap between the unstable region for CWs and the stable region for platicons as the mode coupling increases. We find that while the overlap is negligible when there is no mode coupling, it grows to be a substantial fraction of the platicon stability region as the mode coupling grows and encompasses the zero detuning point. The overlap of the unstable region of the CWs with the stable region of the platicons is a necessary, but not sufficient condition for platicons to self-start. However, we have carried out extensive simulations throughout the region where this necessary condition is fulfilled, and we have found that platicons self-start in every case. Aside from its fundamental interest as an example of symmetry breaking, this result has practical importance since it indicates that it should be possible to obtain microcombs in the normal dispersion regime without too much difficulty as long as the mode coupling is sufficiently large.

\section{The Pump-Split LLE}
\label{sec:linearizationandboundarytracking}  

The version of the LLE that we use is normalized with respect to the linewidth of the pumped cavity resonance $\Delta\omega$, which equals the inverse of the photon lifetime in the cavity, and is written as \cite{coen_modeling_2013,chembo_spatiotemporal_2013,moille_alloptical_2025}
\begin{equation}  \label{eq:bound1}
    \frac{\partial\psi}{\partial t} = -i\sum_mD_\mathrm{int}(\mu)\tilde{\psi}(\mu)\exp{\left(i\mu\theta\right)} + i\lvert\psi\rvert^2\psi - (i\alpha + 1)\psi + F, \tag{1}
\end{equation}
where $\psi$ is the normalized field envelope, $t$ is normalized slow time, $\alpha$ is the normalized detuning, $F$ is the normalized pump power, $D_\text{int}(\mu)$ is the integrated dispersion, $\mu = m - m_0$ is the relative mode number around the mode number of the pumped mode $m_0$. We use $\Tilde{\psi}=\int_{-\pi}^{\pi}\psi(\theta)\exp{\left(-i\mu\theta\right)}({d\theta}/{2\pi})$ to represent the Fourier transform of $\psi$, where $\theta$ is the azimuthal coordinate around the microresonator that rotates at the angular group velocity with respect to a fixed angular coordinate. We will focus on microresonators where third-order and higher dispersion terms can be neglected. Accounting for second-order dispersion and the pump-mode split, we write $D_\mathrm{int}(\mu) = \left(\beta/2\right)\mu^2 - \varepsilon\left[1-\delta(\mu=0)\right]$, where $\beta$ is the normalized second order dispersion, $\varepsilon$ is the normalized mode split strength, and $\delta$ is the Kronecker delta function \cite{lobanov_frequency_2015,yu_continuum_2022}. When $\beta>0$, the dispersion is normal, and when $\varepsilon = 0$, Eq.~\ref{eq:bound1} reduces to the LLE as written in \cite{coen_modeling_2013,chembo_spatiotemporal_2013}. This formulation of a mode split assumes that the physical effect causing the mode split merely disturbs the dispersion of the pumped mode and has no additional effects, allowing us to consider only one set of longitudinal modes.

\section{Platicon Stability Regions}
\label{sec:platiconstabilityregions}

To find the platicon stability regions, we use the dynamical methods developed in \cite{wang_boundary_2014,courtright_automatically_2021,qi_dissipative_2019}. In this approach, we start with a stable stationary solution that we find using a standard evolutionary simulation.  We then continue the solution by changing $\alpha$ or $F$, finding the stationary solution as the parameter changes, and finding in parallel its dynamical spectrum (the spectrum of the linearized evolution operator) to determine its stability.  Once we encounter a stability limit, we track the boundary by continuing the solution just inside the boundary.  We give more details on our approach in the Supplement.

\begin{figure}
    \includegraphics[width=1\linewidth]{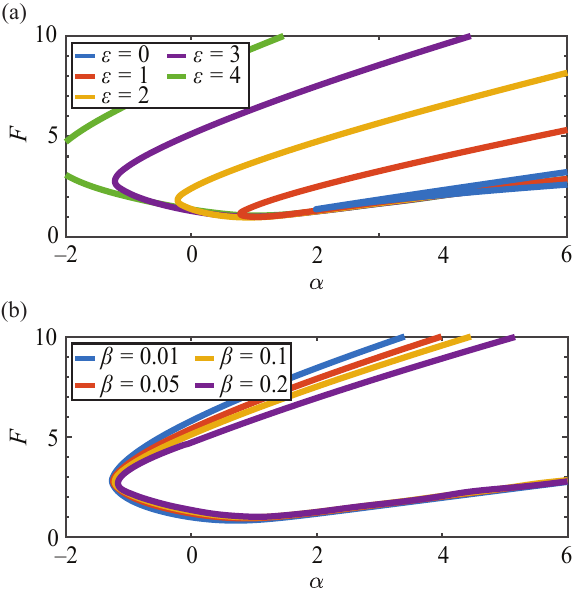}
    \caption{\label{figure1}(a) Stability regions for platicons with a fixed normalized dispersion $\beta = 0.1$ and varied values of the normalized mode split strength $\varepsilon$. (b) Stability regions for platicons with $\varepsilon = 3$ and different values of $\beta$.}
\end{figure}

In Fig.~1, we plot the boundaries of the stability regions for platicons in the $\alpha$-$F$ parameter space where we vary the other two free parameters, the normalized dispersion $\beta$ and the normalized mode split strength $\varepsilon$. Platicon solutions are stable to the right of the curves. From Fig.~1(a), we observe that varying $\varepsilon$ results in large changes to the stability region, with the region growing towards lower detunings and higher powers. Notably, when $\varepsilon=2$, the stability region includes a range of pump powers at zero detuning. This range grows as $\varepsilon$ increases. Conversely, from Fig.~1(b) we observe that changing $\beta$ does not have a significant impact on the stability region. As $\beta$ increases, the lower limit in $F$ shifts slightly downward, while the upper limit shifts upward.  The upper limit demarcates where breathers form when the platicons become unstable.

\begin{figure}[t!]
    \includegraphics[width=1\linewidth]{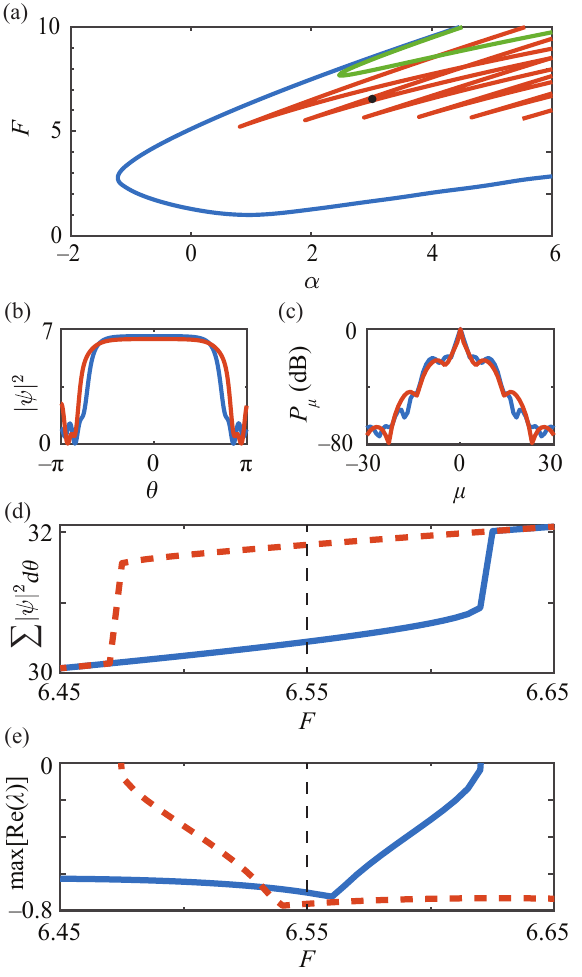}
    \caption{\label{figure1}(a) Stability region for $\varepsilon = 3$ and $\beta = 0.1$. The outer stability boundary is shown in blue. Inner stability boundaries, representing saddle-node bifurcations, are shown in red. Breather solutions exist inside the green curve. Inside each set of two inner stability boundaries that intersect are two stable platicon solutions. The black dot marks the location of the platicons shown in (b) and (c). (b) The intensity profiles of the two stable platicon solutions at $\alpha = 3$ and $F = 6.55$. (c) The spectral profiles of the two stable platicon solutions at $\alpha = 3$ and $F = 6.55$. (d) The intracavity intensity of the available platicon solutions on a scan that either increases (solid blue line) or decreases (dashed red line) $F$ to cross a finger. The dashed black line marks the same $F$ value as the black dot in (a). (e) The largest real eigenvalue for the solutions over the same trajectory as (d). Once the eigenvalue crosses zero, the solution ceases to exist through a saddle-node bifurcation.}
\end{figure}

\begin{figure*}[t]
    \centering
    \includegraphics[width=1\linewidth]{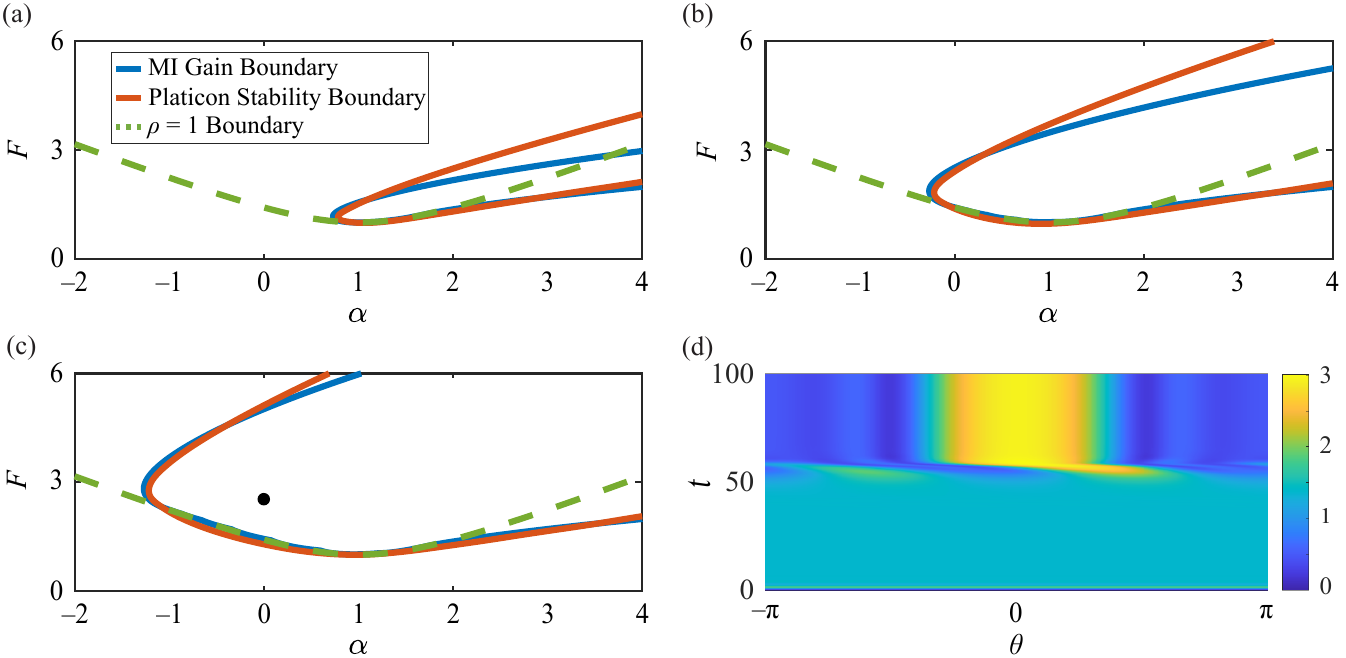}
    \caption{Overlaid MI gain region (blue), platicon linear stability region (red), and CW intensity $\rho = 1$ parabola (green) when $\beta = 0.1$ for different values of $\varepsilon$. (a) $\varepsilon = 1$. (b) $\varepsilon = 2$. (c) $\varepsilon = 3$. The black dot corresponds to the parameters used for the evolution simulation in (d). (d) Intracavity intensity evolution from quantum noise initial conditions at $\alpha = 0$, $F = 2.5$, demonstrating that the platicon self-starts.}
\end{figure*}

In Fig.~2(a), we show additional inner stability boundaries that exist inside the encompassing outer stability boundary for $\varepsilon = 3$ and $\beta = 0.1$. These inner boundary lines form from saddle node bifurcations, and occur when a specific platicon solution trajectory in $\alpha$-$F$ space crosses one of the lines. We see that there are sets of two lines, or ``fingers" that intersect at a point inside the outer stability boundary and grow wider as $\alpha$ and $F$ increase. Inside each of the fingers, there are two distinct stable platicon solutions at each point, leading to bistability. These solutions have different amplitudes in the bright section of the pulse and different numbers of oscillations in the dark section, leading to different spectral profiles. In Figs.~2(b) and (c) we show the intensity and spectra of these bistable solutions at $\alpha = 3$ and $F = 6.55$, marked in Fig.~2(a) by the black dot. Which solution is obtained depends on the $\alpha$-$F$ trajectory taken into and leaving a finger; each solution undergoes a saddle node bifurcation at the other line of the finger from which that solution's trajectory entered. As a result, each solution is stable on a specific side of the finger and both are stable inside. Additionally, we observe a small region of breather solutions near the top of the stability region, the border of which is shown by the green curve in Fig.~2(a). There are regions at large detunings where fingers overlap with each other or the breather region. Within each overlapping area, the solution that is obtained at each $\alpha$-$F$ point depends on how that point is accessed. When two fingers overlap, a third distinct platicon solution appears. As the detuning increases and the number of fingers that overlap increases, the number of solutions that can exist at the same $\alpha$-$F$ point also increases.

In Fig.~2(d), we show the intracavity intensity of the platicon solution inside the cavity as $F$ is first increased and then decreased, moving through the finger at $\alpha=3$ and passing through the point marked by the black dot in Fig.~2(a). In either direction, the platicon remains the same as it enters the finger and then undergoes a sharp transition when leaving the finger to the single stable platicon state. However, the same bifurcation does not appear at the same value of $F$ when the reverse trajectory is traveled; instead the new solution goes through its bifurcation at the other side of the finger. In Fig.~2(e), we show the largest real eigenvalue, or critical eigenvalue, of the same solutions as in Fig.~2(d). We observe that at either end of the finger, one of the solutions undergoes a saddle node bifurcation, represented by the critical eigenvalue moving to zero, after which the bifurcated solution ceases to exist and only a single platicon solution remains.

\section{Platicon Self-Starting}
\label{sec:self-starting} 

In order for a frequency-comb producing waveform to self-start, it is necessary for that waveform to be stable and for an unstable CW mode---due to modulation instability (MI) gain---to exist at the same laser parameters. Self starting was demonstrated by Yang et.~al.~\cite{yang_modulational_2025} using evolutionary simulations in which they showed that platicons can arise spontaneously from noise at specific parameter values with only a frequency perturbation to the pump. This phenomenon was attributed to the expansion of the MI gain region due to the frequency perturbation, but the stability of the platicons beyond those parameter values was not demonstrated.  Here, we use our dynamical approach to determine the region in the $\alpha$-$F$ parameter space in which self-starting is possible. 

In order for a CW mode to be unstable, the intensity of the CW solution $\rho = |\psi_\mathrm{CW}|^2$ must be equal to or greater than 1 \cite{godey_stability_2014}. In $\alpha$-$F$ space, the condition $\rho = 1$ is given by the parabola $F^2 = 1 + (1-\alpha)^2$. CW solutions at $(\alpha,F)$ values above the parabola will always have $\rho > 1$, while those below will have $\rho < 1$ as long as $\alpha < \sqrt{3}$. In the normal dispersion regime without a pump-mode split, an additional phase matching condition stops MI from occurring for CW modes with $\rho > 1$ except in a very small parameter range. The phase matching condition is altered when a pump-mode split is added, such that MI occurs over an extended region proportional to the mode split strength \cite{yang_modulational_2025}. In Fig.~3(a-c), we overlay the extended MI gain regions with the stability regions for $\beta = 0.1$ and $\varepsilon = 1-3$, respectively. We calculated the MI gain regions using equations derived in \cite{godey_stability_2014, yang_modulational_2025}:
\begin{align*}
    F^2 &= \left[1 + (\rho-\alpha)^2\right]\rho, \tag{2.a} \\
    \Gamma(\mu) & = \mathrm{Re}\left\{-1 + \sqrt{\rho^2-\left[\alpha+\varepsilon-2\rho-\frac{1}{2}\beta\mu^2\right]^2}\right\}, \tag{2.b}
\end{align*}
where $\Gamma(\mu)$ is the MI gain parameter, with $\Gamma(\mu) > 0$ indicating that MI will occur.
We observe that the two regions have a large overlap, showing that MI is present in most cases where platicons are stable. When the MI gain region extends past the stability region, unstable CW solutions form but are unable to converge to any platicon solutions. When the stability region extends past the MI gain region, stable platicons solutions exist but cannot be excited from noise. In Fig.~3(d), we show the evolution of the intracavity intensity from quantum noise, with initial conditions with $\beta = 0.1$, $\varepsilon = 3$, $\alpha = 0$, and $F = 2.5$, which corresponds to the black dot in Fig.~3(c). We have examined 50 $(\alpha,F)$ points spread throughout the overlap region where CW solutions are unstable and platicons are stable. We found no cases where the platicons did not self-start.

Below the $\rho = 1$ line when $\alpha > \sqrt{3}$, there are three possible CW solutions with different intensities, as opposed to only one elsewhere \cite{godey_stability_2014}. The intermediate-intensity solution is always unstable and thus inaccessible. Given $\varepsilon> 0$ and a fixed $\alpha$, the highest-intensity solution will always be modulationally unstable over a significant range in $F$ that grows with $\varepsilon$. Conversely, the lowest-intensity CW solution will always be modulationally stable, except for a tiny range of $F$, making it practically impossible to excite. Hence, self-starting is virtually impossible for most of this region despite the modulationally unstable CW upper branch, as the system will naturally converge to the almost entirely stable lower CW branch.

\section{Conclusion}
\label{sec:conclusion} 

In conclusion, we have explored the stability of platicon solutions of the pump-split LLE with different values of mode split strength $\varepsilon$ and dispersion $\beta$. We showed that increasing $\varepsilon$ significantly expands the region where platicons are stable, with the stability region crossing over the zero detuning point when $\varepsilon \geq 2$, while variations to $\beta$ do not affect the stability region strongly. Additionally, we demonstrated the existence of inner stability boundaries inside the overall stable region where certain platicon solutions undergo a saddle-node bifurcation and that the inner stability boundaries combine to form fingers where two distinct platicon solutions are stable at the same pump parameter values, leading to bistability. Finally, we compared our results for stability with the results obtained in \cite{yang_modulational_2025}, where they described the effect of a pump-mode split on the region of MI gain and showed numerically that platicons can self-start. We showed that the MI gain and linear stability regions for platicons largely overlap for standard $\alpha$ and $F$ parameters in the absence of thermal effects, demonstrating that platicon self-starting is possible over a broad parameter range that grows as $\varepsilon$ grows. Additionally, the overlapping region moves to cover zero detuning once $\varepsilon \geq 2$, indicating that self-starting platicons can be generated by pumping on or near resonance instead of only with significant detuning.

\section{ACKNOWLEDGEMENTS}
L.C., A.S., P.S., and C.M. acknowledge support from The National Center for Manufacturing Sciences (Cooperative Agreements 2022138-142232 and 2023200-142386 as subawards from DoD Cooperative Agreements HQ0034-20-2-0007 and HQ0034-24-2-0001) and from the National Institute of Standards and Technology (Grant No.~60NANB24D106).  L.C. also acknowledges support from the DoD SMART program.

\section{DATA AVAILABILITY}
The data that support the findings of this article are publicly available (to be uploaded).

\section{AUTHOR CONTRIBUTIONS}
L.C. and C.R.M. conceived the project. L.C. carried out the  simulations. L.C. and C.R.M. wrote the first draft of the manuscript. All authors reviewed and edited the manuscript.

\bibliography{PlaticonPaper}

\clearpage

\onecolumngrid

\begin{center}
    {\Large\bfseries Supplemental Material\par}
    \vspace{1em}

    {\large\bfseries
    Platicon Stability Regions, Bistable Solutions, and Self-Starting
    in Photonic Crystal Microresonators\par}
    \vspace{1em}

    Logan Courtright, Alexey Sukhinin, Pradyoth Shandilya,\\
    Tanvir Mahmood, James P. Cahill, and Curtis R. Menyuk
\end{center}

\vspace{2em}

\twocolumngrid

\section{Linear Stability Analysis}

To study the stability regions of platicons, we use the dynamical methods developed in \cite{wang_boundary_2014,courtright_automatically_2021,qi_dissipative_2019} to study the linear stability of the pump-split Lugiato-Lefever Equation (LLE). For ease of computation,  we combine the Taylor-series derivative terms with the pump-mode split term to create a general dispersion operator $\mathcal{D}_m$, written as
\begin{equation} \label{eq:bound1}
    \mathcal{D}_m = \mathcal{F}t^{-1}\left[\{\left(\beta/2\right)\mu^2 - \varepsilon\left[1-\delta(\mu=0)\right]\}\Tilde{\psi}\right],
\end{equation}
where $\beta$ is the normalized second order dispersion, $\varepsilon$ is the normalized mode split strength, and  $\Tilde{\psi}$ is the Fourier transform of $\psi$.

To start our study, we find a stationary solution $\psi_0 = \psi_{r0}+i\psi_{i0}$ of Eq.~\textcolor{blue}{1} in the main text, where $\psi_{r0}$ and $\psi_{i0}$ are the real and imaginary parts of $\psi_0$ respectively. We do so by using a variant of the Newton-Raphson method applied to Eq.~\textcolor{blue}{1} with $\partial\psi/\partial t$ set to zero. We then linearize the pump-split LLE around the real $\left(\psi_{r0}\right)$ and imaginary $\left(\psi_{i0}\right)$ parts of $\psi_0$ such that
\begin{equation} \label{eq:bound2}
 \begin{split}
    \psi_r\left(\theta,t\right) &= \psi_{r0}\left(\theta\right) + \Delta\psi_r\left(\theta,t\right) \\
	\psi_i\left(\theta,t\right) &= \psi_{i0}\left(\theta\right) + \Delta\psi_i\left(\theta,t\right),
     \end{split}
\end{equation} 
where $\Delta\psi_r$ and $\Delta\psi_i$ are small perturbations to the real and imaginary parts of $\psi_0$. The perturbations $\Delta\psi_r$ and $\Delta\psi_i$ obey the linearized equation
\begin{equation} \label{eq:bound3}
	\pypx{\Delta\bm{\Psi}}{t} = \mathcal{L}\Delta\bm{\Psi},
\end{equation}
where
\begin{align}		
	\mathcal{L} &= 
	\begin{bmatrix}
           -2\psi_r\psi_i-1+\mathcal{D}_{m,i} & -\psi_r^2-3\psi_i^2+\alpha+\mathcal{D}_{m,r}\\
		3\psi_r^2+\psi_i^2-\alpha+\mathcal{D}_{m,r} & 2\psi_r\psi_i-1+\mathcal{D}_{m,i}
	\end{bmatrix},
\end{align}
and
\begin{align}		
	\Delta\bm{\Psi} &= 
	\begin{bmatrix}
		\Delta\psi_i\\
		\Delta\psi_r
	\end{bmatrix}.
\end{align}
The parameters $\mathcal{D}_{m,r}$ and $\mathcal{D}_{m,i}$ are the real and imaginary parts of $\mathcal{D}_m$ respectively.
The eigenvalues of the discretized matrix $\mathcal{L}$ in Eq.~\ref{eq:bound3} determine the stability of the platicon solution about which we perturb. If any of the eigenvalues are positive, the system is unstable; therefore, we check that all the eigenvalues are negative except for one eigenvalue that is fixed at the origin due to translational invariance of Eq.~(1) in the main text. From here, we systematically continue the solution through the $\alpha$-$F$ parameter space checking for stable solutions, which allows us to map out a region where stable solutions exist. As we vary $\alpha$ and $F$, the platicon solution  eventually becomes unstable via a bifurcation that changes the solution; different bifurcations types lead to different outcomes. The outcome can be a new stationary solution, a continuous wave, a breather, chaos, or some other type of evolving solution. Once we have found a boundary for a stable region, we continue the solution using three trajectories just inside the boundary while extrapolating to the boundary. We refer to thfis procedure as boundary tracking. Cusps occur on the boundary where the bifurcation type changes, requiring special treatment. More details on this procedure can be found in Ref.~\cite{courtright_automatically_2021}.

\end{document}